\documentstyle[12pt]{article}

\begin{document}
\newcommand{\ta}{\theta^{1,0\;\und i}}
\newcommand{\tb}{\theta^{1,0}_{\und i}}
\newcommand{\tc}{\theta^{0,1\;\und a}}
\newcommand{\td}{\theta^{0,1}_{\und a}}
\newcommand{\da}{D^{2,0}}
\newcommand{\db}{D^{0,2}}
\newcommand{\du}{D^{0,0}_u}
\newcommand{\dv}{D^{0,0}_v}
\newcommand{\na}{N^{2,0}}
\newcommand{\nb}{N^{0,2}}
\newcommand{\nn}{\nonumber}
\newcommand{\be}{\begin{equation}}
\newcommand{\bea}{\begin{eqnarray}}
\newcommand{\eea}{\end{eqnarray}}
\newcommand{\ee}{\end{equation}}
\newcommand{\eps}{\varepsilon}
\newcommand{\und}{\underline}
\newcommand{\p}[1]{(\ref{#1})}

\begin{titlepage}
\begin{flushright}
ENSLAPP-L-588\\ 
JINR E2-96-155 \\
hep-th/9604186 \\
April 1996
\end{flushright}
\vskip 1.0truecm
\begin{center}
{\large \bf MORE ON $(4,4)$ SUPERMULTIPLETS 
IN $SU(2)\times SU(2)$ HARMONIC SUPERSPACE}
\vglue 2cm
{\bf E. Ivanov${}^{(a, b)}$ and A. Sutulin${}^{(b)}$}
\vglue 1cm 
{${}^{(a)}$ \it Lab. de Phys. Th\'eor. ENSLAPP, ENS -Lyon} \\
{\it 46 All\'ee d'Italie, 69364 Lyon, France}
\vglue 3mm
{${}^{(b)}$ \it Bogoliubov Laboratory of Theoretical Physics, 
JINR,\\ 
141 980 Dubna, Moscow Region, Russia}
\end{center}

\vspace{2cm}

\begin{abstract}
We define the $SU(2)\times SU(2)$ harmonic superspace analogs of 
tensor and nonlinear $(4,4)$, $2D$ supermultiplets. They are described 
by constrained analytic superfields and provide an off-shell 
formulation of a class of torsionful $(4,4)$ supersymmetric sigma 
models with abelian translational isometries on the bosonic 
target. We examine their relation to $(4,4)$ twisted 
multiplets and discuss different types of $(4,4)$ dualities 
associated with them. One of these dualities implies the standard abelian 
$T$-duality relations between the bosonic targets in the initial 
and dual sigma model actions. We show that $N=4$, $2D$ superconformal  
group admits a simple realization on the superfields introduced, and 
present a new superfield form of the $(4,4)$ $SU(2)\times U(1)$ WZNW action.  
\end{abstract}

\end{titlepage}

\section{Introduction}
 
For unambiguous construction of models with $2D$ supersymmetry, 
in particular, supersymmetric string models, it is of crucial importance 
to have the full list of off-shell representations of 
$2D$ supersymmetry, as well as to know various interrelations between them, 
e.g., via duality transformations. Working with 
off-shell supermultiplets, especially in the superfield approach, allows 
one to keep supersymmetry manifest at each step and provides simple general 
rules of the model-building. 

Two important off-shell multiplets of $N=2$, $4D$ supersymmetry are the 
tensor \cite{ten1} and nonlinear \cite{a1} ones. They were primarily  
used as compensators breaking $N=2$ conformal supergravity down to some off-shell versions of Einstein $N=2$ supergravity \cite{a1}. 
Later on, they were exploited to construct a subclass of $N=2$, $4D$ 
supersymmetric sigma models and 
to explicitly compute the relevant bosonic 
hyper-K\"ahler metrics \cite{{a2},{a3}}. 
Upon reduction $N=2$, $4D$ $\rightarrow$ $N=(4,4)$, $2D$, such models 
can provide some string backgrounds, and this is the main reason of recent 
revival of interest to these multiplets (and some their further 
generalizations) in the context of string theory 
\cite{{string1},{string2},\cite{a4}}. 
In particular, in \cite{a4} it has been proposed to utilize the nonlinear multiplet from this point of view. 

Sigma models associated with these multiplets 
yield no 
torsion in the bosonic part of the action, the relevant target manifolds 
are hyper-K\"ahler \footnote{To be more precise, this is entirely true 
only in the dual representation of the relevant action through 
hypermultiplets, see Sect. 2.}. 
On the other hand, generic string backgrounds possess a nontrivial torsion. 
The basic aim of 
the present paper is to propose a generalization of these $(4,4)$ multiplets, 
such that the relevant sigma models actions contain the torsion terms which 
cannot be removed by any duality transformation with preserving 
manifest $(4,4)$ supersymmetry. 

The natural off-shell description of torsionless $(4,4)$ supersymmetric 
sigma models is achieved within the 
$2D$ version of $SU(2)$ harmonic 
superspace (HSS) \cite{HSS}. In refs. \cite{{tens},{nonl},{obzor}} 
the tensor and 
nonlinear multiplets were formulated as $SU(2)$ harmonic analytic 
superfields with a restricted dependence on $SU(2)$ harmonics. 
As follows from the 
results of ref. \cite{dual}, general off-shell interactions of the 
tensor and some other $(4,4)$ multiplets with finite sets of 
auxiliary fields are equivalent, 
via a superfield duality transformation, to some particular classes of 
self-interaction of the ultimate off-shell $(4,4)$ hypermultiplet, the 
unconstrained harmonic analytic superfield $q^{+}$ with infinite number of 
auxiliary fields. Though the case of nonlinear multiplet was missed in 
ref. \cite{dual}, the previous statement is true for it as well, and 
in Sect. 2 we will demonstrate this.
    
As was argued in refs. \cite{{IS},{EI},{EI1}}, the appropriate framework 
for off-shell 
description of $(4,4)$ supersymmetric sigma models with torsion 
is provided by a generalization of $SU(2)$ HSS, viz., the doubly extended 
$SU(2)\times SU(2)$ harmonic superspace \footnote{$SU(2)\times SU(2)$ HSS is akin to the $(4,4)$ projective superspace which was earlier introduced in \cite{R1}.}. So, in order 
to generalize the 
tensor and nonlinear multiplets to the case with torsion it is natural to 
look for the $SU(2)\times SU(2)$ HSS analogs of the superfields 
by which these multiplets are represented in the $SU(2)$ HSS.
  
We start in Sect. 2 by recalling the formulation 
of tensor and nonlinear $(4,4)$ supermultiplets in $SU(2)$ HSS. Then, 
in Sect. 3, we generalize to $SU(2)\times SU(2)$ HSS the 
defining constraints of these multiplets in $SU(2)$ HSS. The relevant 
$SU(2)\times SU(2)$ harmonic analytic superfields propagate $(16+16)$ 
physical fields, as distinct from their $SU(2)$ harmonic prototypes which 
propagate $(4+4)$ such fields. We demonstrate that general 
self-interactions of these new superfields are off-shell 
equivalent to particular classes of self-interactions of four 
twisted chiral $(4,4)$ multiplets. These subclasses are distinguished 
in that they possess abelian translational isometries. The superfields 
defined here seem to be most appropriate for describing this type of 
torsionful $(4,4)$ sigma models. In Sect. 4 we discuss some 
peculiarities of the relevant superfield sigma model 
actions. In particular, we point out the existence of different  
dual formulations of them. In one of these formulations the dual bosonic lagrangian is related to the original one by the standard $T$ duality 
relations \cite{Buscher}. We also present the realization of the 
world-sheet $N=4$, $SU(2)$ superconformal group on the superfields 
introduced, as well as a new superfield form of the $(4,4)$ supersymmetric $SU(2)\times U(1)$ WZNW action.   
 
\setcounter{equation}{0}

\section{Tensor and nonlinear multiplets in $SU(2)$ \\ 
harmonic superspace}
In the $2D$ version of $SU(2)$ HSS approach \cite{HSS} 
the tensor multiplet is represented by the superfield $L^{(++)}$
which (i) lives on the harmonic analytic $(4,4)$, $2D$ superspace
\be  \label{analytsu2}
\{\zeta^M, u \} \equiv \{ x^{\pm \pm}_a, \theta^{(+)\;\pm}, 
\bar \theta^{(+)\;\pm}, u^{(+)i}, u^{(-)j} \}, \; \;L^{(++)} \equiv L^{(++)}(\zeta, u)\;, \ee
(ii) is real 
\be  \label{realityL}
L^{(++)} = \widetilde{L}^{(++)}
\ee
and (iii) obeys the following constraint \cite{tens}
\be
D^{(++)}L^{(++)} = 0\;. 
\label{tensor}
\ee 
In these formulas, the indices $\pm$ without and with parentheses 
are, respectively, the $2D$ Lorentz and harmonic $U(1)$ charge ones 
(this $U(1)$ charge is assumed to be strictly preserved), the quantities 
$u^{(+)i},\; u^{(-)i}$,
$$
u^{(+)i}u^{(-)}_i = 1, \;\; 
u^{(+)i}u^{(-)k} - u^{(+)k} u^{(-)i} = -\epsilon^{ik},
$$
are harmonic variables parametrizing the group $SU(2)_A$, one of the 
diagonal $SU(2)$'s in the full $(4,4)$ supersymmetry automorphism group 
$SO(4)_L\times SO(4)_R$, the symbol $\widetilde{}$ means a generalized 
involution with respect to which the 
superspace \p{analytsu2} is real, and $D^{(++)}$ is the 
analyticity-preserving harmonic derivative 
\be
D^{(++)} = u^{(+)i}\frac{\partial}{\partial u^{(-)i}} +
i \theta^{(+)+} \bar \theta^{(+)+} \partial_{++}  
+ i \theta^{(+)-} \bar \theta^{(+)-} \partial_{--}\;.
\ee
More details of the harmonic superspace approach can be found, e.g., in 
refs. \cite{HSS} - \cite{obzor}. Recall that unconstrained analytic 
harmonic superfields contain an infinite tail of auxiliary fields 
arising from the harmonic expansion on the two-sphere 
$S^2 \sim SU(2)_A/U(1)_A$ (the expansion just on $S^2$ instead of 
the whole group $SU(2)_A$ comes out as the result of the preservation of harmonic $U(1)$ charge in the harmonic superspace formalism). 
The role of the constraint \p{tensor} is to reduce this infinite 
tail to the standard $(8+8)$ off-shell component content 
of the tensor multiplet. 

The characteristic feature of the superfield $L^{(++)}$ is that one 
of its physical bosonic degrees of freedom is supplied by the 
$2D$ vector $V_{\pm \pm}$ (``notoph'') subjected to the constraint 
\be
\partial_{++} V_{--} + \partial_{--} V_{++}= 0 \label{linV}
\ee 
that is implied by the superfield one \p{tensor} (the component fields 
$V_{\pm\pm}$ enter $L^{(++)}$ as the coefficients of the 
$\theta$ monomials $\theta^{(+)+}\bar \theta^{(+)+}$, 
$\theta^{(+)-}\bar \theta^{(+)-}$, respectively). Eq. \p{linV} 
can be solved as  
\be \label{solL}
V_{\pm\pm} = \pm i\partial_{\pm \pm} \phi(x), \;\; \left( \phi^{\dagger} = 
\phi \right)\;, 
\ee
thus introducing the fourth bosonic scalar field.

The general $L^{(++)}$ action reads \cite{tens}
\be 
S_L = {1\over \kappa^2} \int \mu^{(-4)} 
\tilde{F}^{(+4)} (u, L^{(++)})\;. \label{Laction}
\ee
Here $\tilde{F}^{(+4)}$ is an arbitary function of its arguments with 
the appropriate flat part 
$$
\tilde{F}^{(+4)} = - L^{(++)}L^{(++)} + O(L^2)\;, 
$$
$\kappa$ is the dimensionless sigma model coupling constant and 
$\mu^{(-4)}$ is the analytic superspace integration measure
$$
\mu^{(-4)} = d^2 x_a d^2 \theta^{(+)+} d^2 \theta^{(+)-} [du], 
$$
($[du]$ denotes the integration over two-sphere $S^2$).
An extension to the case of several $L^{(++)}$ is obvious.

The general distinguishing property of this action is the abelian 
translational isometry 
realized as a shift of the $SU(2)$-singlet field $\phi(x)$ coming out as the 
solution to the notoph constraint \p{linV}; as the result, the corresponding 
bosonic metrics do not depend on this field. The constraint \p{tensor} can be implemented in the action with the help of 
the analytic superfield lagrange multiplier $\omega$ to yield a dual 
$\omega$ formulation of the action \cite{dual} 
\be
S_{L,\omega} = S_L + {1\over \kappa^2} \int d\zeta^{(-4)} \;\omega \; D^{(++)}L^{(++)} \;. 
\label{Ldual}
\ee
It also possesses an $U(1)$ isometry, 
this time realized as shifts of $\omega$. The dual $\omega$ 
form of the general $L^{(++)}$ action \p{Laction} can be obtained by 
eliminating $L^{(++)}$ by its algebraic equation of motion. The dual 
action yields in the bosonic 
sector the most general four-dimensional hyper-K\"ahler metric with one 
translation isometry (in the case of $n$ copies of $L^{(++)}$, the most 
general $4n$ dimensional hyper-K\"ahler metric with $n$ mutually 
commuting $U(1)$ isometries) \cite{obzor}. The action \p{Laction} 
in its own right produces, along with the metric, also a non-zero 
torsion; these both are related to the dual hyper-K\"ahler metric 
by the well-known Buscher's formulas \cite{Buscher}. 

Duality transformation in HSS has been firstly 
introduced in \cite{HSS2} and later on has been used to show that the 
off-shell 
actions of various matter multiplets of $N=2$, $4D$ ($(4,4)$, $2D$) 
supersymmetry 
with finite numbers of auxiliary fields are duality-equivalent to 
particular classes of the general action of the analytic 
$q^{(+)}$ hypermultiplet with an infinite number of auxiliary 
fields \cite{{tens},{dual}}. 
The basic feature of this kind of duality transformation 
is the preservation of manifest $N=2$ supersymmetry (or 
$(4,4)$ supersymmetry in the two-dimensional case) 
at each step. 

Combining the superfields $\omega$ and $L^{(++)}$ into the single 
unconstrained analytic superfield $q^{(+)i}$
\be
q^{(+)i} \equiv u^{(-)i} L^{(++)} - {1\over 2} u^{(+)i} \omega\;, 
\;\;\; 
L^{(++)} = u^{(+)i} q^{(+)}_i,\; \omega = 2 u^{(-)i} q^{(+)}_i\;, 
\label{qLom}
\ee
where we have made use of the property of completeness of the harmonics, 
the action \p{Ldual} can be indeed rewritten as a particular 
representative of actions of the superfield $q^{(+)i}$. This 
superfield is ``ultimate'' for torsionless $(4,4)$ sigma models, 
in the sense that its most general 
self-interactions yield most general hyper-K\"ahler metrics in the 
bosonic sector \cite{HK}. This way, the theorem about the 
relationship between $(4,4)$, $2D$ worldsheet supersymmetry 
($N=2$ in four dimensions) and bosonic target manifolds of the related 
torsionless sigma models \cite{AGF} is visualized. General actions 
of $q^{(+)}$ possess no any isometries and do not admit a 
duality transformation to the form with a finite number of 
auxiliary fields.

Let us turn to discussing the nonlinear multiplet. As no a 
systematic treatment of it has been given so far in the literature 
on the HSS approach, we dwell on this subject in some more detail. 
 
In the $SU(2)$ $2D$ HSS the nonlinear multiplet 
is described by the real analytic superfield $N^{(++)}(\zeta, u)$ 
subjected to the constraint \cite{{nonl},{obzor}}
\be
D^{(++)}N^{(++)} + (N^{(++)})^2 = 0\;. 
\label{harmconstrN}
\ee
Once again, the role of this constraint is to reduce an infinite 
tail of the fields appearing in the harmonic decomposition 
with respect to the variables $u$ to the standard 
off-shell component content of nonlinear multiplet which is $(8+8)$ as 
in the case of tensor multiplet. The equivalence of 
the analytic superspace description of the nonlinear multiplet 
to the one in the conventional $(4,4)$, $2D$ superspace 
\cite{a1} can be easily demonstrated \cite{{quat},{obzor}}. 
 
Taking into account the constraint \p{harmconstrN}, one can show that 
the most general action for $k$ independent superfields 
$N^{(++)}_\alpha \ (\alpha=1,\ldots,k)$ reads
\be
S_N^k = {1\over \kappa'^2}\int \mu^{(-4)} F^{(+4)}_N(N^{(++)}_1, 
\ldots,N^{(++)}_k, u)\;.
\label{Naction}
\ee
The action is particularly simple for just one $N^{(++)}$:
\be
S_N^1 = {1\over \kappa'^2}\int d\zeta^{(-4)} 
N^{(++)}(\zeta)c^{(++)}(u^\pm),
\label{Naction1}\ee
where $c^{(++)}$ is an arbitrary function of the harmonics. 
Any power of $N^{(++)}$ can be reduced to a term linear 
in $N^{(++)}$ with making use of the constraint (\ref{harmconstrN}) 
and integrating by parts with respect to harmonic variables 
(harmonic integrals of $D^{++}$ applied on anything vanish). 

We wish to point out that the actions for 
$N^{(++)}$ involve explicitly only 3 out of the 4 physical scalars of 
the on-shell matter multiplet. The fourth scalar, as in the case of  
tensor multiplet, is supplied by
the constrained vector field $V_{\pm \pm}(x)$ ($N^{(++)} = 
i\theta^{(+)+} \bar \theta^{(+)+}\;V_{++}(x) + 
i\theta^{(+)-} \bar \theta^{(+)-}\; V_{--}(x) +  \ldots$). This 
constraint follows from (\ref{harmconstrN}):
\be
\partial_{++} V_{--} + \partial_{--} V_{++} + 2\;V_{++} V_{--} = 0
\label{nonlV}
\ee
(neglecting contributions from other fields). 
Unlike the notoph constraint \p{linV},
eq. (\ref{nonlV}) cannot be solved
explicitly. It seems that the only reasonable way to deal 
with (\ref{nonlV}) is to implement it in the action with a scalar 
Lagrange multiplier. 
The latter becomes the fourth bosonic degree of freedom upon 
elimination of $V_{\pm \pm}$ and the action of the nonlinear multiplet 
acquires the standard sigma model form. This
naturally comes about within the dual description of 
nonlinear multiplet in terms of unconstrained analytic superfields. 

In \cite{dual} the case of nonlinear multiplet was missed. 
Here we fill this gap. For simplicity we will consider the case 
of one $N^{(++)}$.   

To obtain the dual action in this case, we insert the 
constraint \p{harmconstrN} 
into \p{Naction1} with the help of suitable Lagrange multiplier:
\be
S_N^1 = {1\over \kappa'^2}\int d\zeta^{(-4)}\left\{ 
N^{(++)}(\zeta, u)c^{(++)}(u) + \omega \left[D^{(++)}N^{(+)} + 
(N^{(++)})^2 
\right] \right\}\;.
\label{dualN1}\ee
Varying this action with respect to  $\omega$, we come back to the 
constraint \p{harmconstrN} and action \p{Naction1}. On the other hand, 
varying with respect to $N^{(++)}$, we get 
\be  \label{Neqalg}
2N^{(++)}\omega = D^{(++)}\omega - c^{(++)}.
\ee
Assuming that $\omega$ starts with a constant (i.e. that we can divide by 
$\omega= 1 + ... $), redefining it as 
\be  \label{redefom}
\omega = \hat \omega^2
\ee
(this is a canonical redefinition in virtue of the previous 
assumption) and substituting all this back into \p{dualN1}, we obtain the 
dual $\omega$ representation of the latter in the following form  
\be
S^{\mbox{\scriptsize dual}}_{\omega} = 
{1\over \kappa'^2}\int \mu^{(-4)} \left[
-(D^{(++)}\hat\omega)^2 -{1\over 4}{(c^{(++)})^2 
\over \hat\omega^2}
- \ln \hat \omega \;D^{(++)}c^{(++)}\right].
\label{omega1}\ee

In the particular case
\be  \label{EHconstr}
D^{(++)}c^{(++)} = 0\;, \Rightarrow c^{(++)} = 
c^{(ik)}u^{(+)}_iu^{(+)}_k\;, 
\ee 
the action \p{omega1} 
takes a form which highly resembles the $\omega$ representation of 
the action of $(4,4)$ sigma 
model with the Eguchi-Hanson manifold as the bosonic target 
\cite{Town}
\be
S^{\mbox{\scriptsize dual}}_{\omega,N} = {1\over \kappa^2}
\int \mu^{(-4)} \left[
-(D^{(++)}\tilde\omega)^2 -{1\over 4}{(c^{(++)})^2 
\over \tilde\omega^2}\right].
\label{EHwrong}\ee
The only difference is in the sign of the second term, so in 
the present case we obtain the EH metric with the wrong sign of 
the ``mass''-parameter (which is just $(c^{12})^2$ in the 
fixed $SU(2)_A$ frame). The $N=(2,2)$ superfield form of 
the same action has been given in \cite{a3}. Note that the 
invariance group of the action \p{EHwrong} 
is $SU(2)\times U(1)$, just as in the case of 
the standard EH $(4,4)$ sigma model \cite{Town}, $SU(2)$ being a kind 
of Pauli-G\"ursey group commuting with $(4,4)$ supersymmetry while 
$U(1)$ a part of the $(4,4)$ supersymmetry automorphism group. 
On the bosonic target manifold this $SU(2)\times U(1)$ is realized as 
isometries of the target metric, respectively as 
``translational'' and ``rotational'' ones, in agreement with 
the fact that the EH metric possesses such isometries 
\cite{reviewHK}. The explicit realization of the 
$SU(2)$ factor of the isometry group in the action \p{EHwrong} 
coincides with that given in ref. \cite{Town}. 
Of course, the original $N^{(++)}$ action \p{Naction1} with 
the specific $c^{(++)}$ \p{EHconstr} and constraint \p{harmconstrN} 
respects the same invariance group properly realized on $N^{(++)}$.

Note that it is easy to rewrite the $\omega, N^{(++)}$ action 
\p{dualN1} as a subclass of general actions of the ultimate analytic 
$q^{(+)}$ hypermultiplet, in accord with the statement that the general $q^{(+)}$ 
action corresponds to the most general hyper-K\"ahler off-shell $(4,4)$ 
supersymmetric sigma model. 
The original $\omega, N^{(++)}$ representation (as well as 
the $\omega, L^{(++)}$ representation of the tensor multiplet 
action) turns out to be more preferrable for generalizing to 
the case with torsion. 

Finally, we make two comments. Firstly, as  
is seen from \p{omega1}, even in the case when the original action 
\p{Naction1} is zero ($c^{(++)} = 0$), its dual \p{omega1} is non-trivial 
and describes a free hypermultiplet. This subtlety has been also noticed 
and discussed in \cite{a4} in the framework of $(2,2)$ superfield 
formalism. It can be traced to the above-mentioned fact that the actions \p{Naction}, \p{Naction1} as they stand admit no standard sigma model interpretation which becomes possible only after passing to the 
dual description (the assumption that the lagrange multiplier superfield $\omega$ contains a non-zero ``classical'' constant part is important for self-consistency of such a description). 

Another comment concerns the relation to the $(4,4)$, $2D$ 
tensor multiplet superfield $L^{(++)}$. Its defining constraint 
\p{tensor} can be regarded as a degenerate 
limit of \p{harmconstrN} (one rescales $N^{(++)}= \gamma L^{(++)}$, 
substitutes this into \p{harmconstrN}, divide by $\gamma$ and finally 
put $\gamma$ equal to zero), however the actions of $L^{(++)}$ are 
radically differ from those of $N^{(++)}$ and cannot be related 
to the latter by any limiting procedure. 
In contrast to the case of nonlinear multiplet, in the dual action 
of $L^{(++)}$ \p{Ldual}
the lagrange multiplier term alone (with $\tilde{F}^{(+4)}= 0$) 
produces no non-trivial action: varying 
$L^{(++)}$ yields $D^{(++)}\omega = 0$ $\Rightarrow \omega = const$.  

\setcounter{equation}{0}

\section{Generalizations to the $SU(2)\times SU(2)$ harmonic \\
superspace}

The $(4,4)$ $SU(2)\times SU(2)$ HSS is an extension of the
standard real $(4,4)$ $2D$ superspace by two independent sets
of harmonic variables $u_i^{\pm1}$ and $v_a^{\pm1}$ associated
with two commuting automorphism groups $SU(2)_L$ and $SU(2)_R$
of the left and right sectors of $(4,4)$ supersymmetry 
\cite{{IS},{R1}}. The $SU(2)\times SU(2)$ HSS formalism enables one to 
keep both these $SU(2)$ symmetries manifest at each step and to control 
their breakdown.

In what follows we will be interested in an analytic 
subspace of the $SU(2)\times SU(2)$ HSS. It is
presented by the following set of coordinates 
\be
(\zeta, u, v)= (x^{++}, x^{--},\theta^{1,0\,\und i},
\theta^{0,1\,\und a},u^{\pm1\,i},v^{\pm1\,a}), \label{anal}
\ee
and is closed under the $(4,4)$ supersymmetry transformations.
The pairs of superscripts $``n,m''$ in \p{anal} stand for the values 
of two independent harmonic $U(1)$ charges which, like in the case 
of $SU(2)$ HSS, are assumed to be strictly 
conserved. As the result of this requirement, all superfields defined 
on \p{anal}, the $SU(2)\times SU(2)$ analytic $(4,4)$ superfields, 
are expanded in the double harmonic series on the product 
$SU(2)_L/U(1)_L \otimes SU(2)_R/U(1)_R$. 
Extra doublet indices 
$\und i, \und a$ of Grassmann coordinates in \p{anal} refer to 
two additional $SU(2)$ automorphism groups of $(4,4)$ supersymmetry 
which, together with $SU(2)_L$ and $SU(2)_R$, constitute the full 
automorphism group $SO(4)_L\times SO(4)_R$ of the latter. We omit 
the $2D$ Lorentz indices of Grassmann coordinates, keeping in mind 
that the first and second $\theta$'s in \p{anal} carry, respectively, 
the indices $+$ and $-$. 

In the present case one can define two harmonic 
derivatives preserving the analyticity, the left and right ones 
\be  \label{deriv}
\da = \partial^{2,0} + i \ta \tb \partial_{++} \;, \;\;\;\;\;\;
\db = \partial^{0,2} + i \tc \td \partial_{--}\;.
\ee
Their very important property having no analogs in the $SU(2)$ HSS 
case is their commutativity  
\be
[\da, \db] = 0\;. \label{commut}
\ee
As we will see, it places severe restrictions on the possible form 
of the constraints one can impose on the $S(2)\times SU(2)$ analytic 
superfields in order to cut an infinite tail of auxiliary fields in 
their $u,v$ harmonic expansions and thus to get $(4,4)$ multiplets 
with finite sets of fields. 

Our further aim will be to discuss possible generalizations of 
the $SU(2)$ harmonic constraints  \p{tensor}, \p{harmconstrN}
to the $SU(2)\times SU(2)$ case. 
The natural primary requirements are 
(i) these constraints involve first degrees of $\da, \db$ ; (ii) 
they do not give rise to any dynamical equation for the component 
fields, i.e. are purely kinematic. 

We start with discussing $SU(2)\times SU(2)$ analogs of the 
linear constraint \p{tensor} as the simplest one. One of such sets  
has been already presented in \cite{IS}, it is the constraints 
defining $(4,4)$ twisted multiplet 
\be 
\da q^{1,1} = \db q^{1,1} = 0 \;, \label{constrq}
\ee
where $q^{1,1} (\zeta,u,v)$ is an analytic superfield. Like 
\p{tensor} in application to $L^{++}$, they leave in $q^{1,1}$ 
$(8+8)$ independent components, including $4$ physical boson fields. 
However, as was noticed in \cite{{IS},{EI}}, the mechanism of 
achieving this irreducible content is different 
for \p{tensor} and \p{constrq}. 
While the fourth bosonic field in $L^{(++)}$ is supplied by a 
divergenceless $2D$ vector, \p{constrq} amount to 
purely algebraic relations between the components of $q^{1,1}$, so 
that all four physical bosons appear on equal footing 
as the components of the $4\times 4$ matrix $q^{ia}(x)$, 
$q^{1,1} = q^{ia}(x) u^1_iv^1_a + ...$ . No any constrained 
vectors are present. Also it is easy to see that \p{constrq} 
do not admit a nonlinear extension like \p{harmconstrN}. Indeed, 
without allowing for extra harmonic charged constants it is impossible 
to construct nonlinear addings to the l.h.s. of eqs. \p{constrq} out of 
$q^{1,1}$, so that they possess the harmonic charges $(3,1)$ 
and $(1,3)$ \footnote{Even with such constants included, any 
nonlinear modification of \p{constrq} is reduced to \p{constrq} 
via a canonical redefinition of $q^{1,1}$ \cite{{EI},{EI1}}. 
This is a consequence 
of the commutativity condition \p{commut}.}. For further reference we 
present the most general action of $n$ copies of $q^{1,1}$ multiplet 
\be 
S_{q^{1,1}} = \int \mu^{-2,-2} L^{2,2} (q^{1,1\;M}, u,v)\;,  \; 
\mbox{det} \frac{\partial^2 L^{2,2}}{\partial q^{1,1\;M} \partial q^{1,1\;N}}|_{q^{1,1} = 0} \neq 0\;\; (M, N =1,...n)\;. 
\label{q11action}
\ee
Here $\mu^{-2,-2} = d^2xd^2\theta^{1,0}d^2\theta^{0,1}[du][dv]$ is the 
analytic superspace integration measure.

As another possible generalization of \p{tensor} which was not discussed 
so far, we introduce two 
$SU(2)\times SU(2)$ analytic superfields $q^{2,0}$, $q^{0,2}$ subjected 
to the constraints 
\be 
\da q^{2,0} = 0\;, \;\; \db q^{0,2} = 0 \;. \label{q1}
\ee
The appearance of just two superfields is necessary in order to be 
able to construct the relevant free action which is given by 
\be 
S_{free} \propto \int \mu^{-2,-2} q^{2,0} q^{0,2}\;. \label{q20act}
\ee
No any meaningful action can be constructed out of $q^{2,0}$ or 
$q^{0,2}$ alone.

The constraints \p{q1} do not restrict the $v$ dependence in $q^{2,0}$ 
and the $u$ dependence in $q^{0,2}$. Besides, they put no any relation 
between these superfields. If one solves \p{q1} and 
substitutes the solution into  \p{q20act}, no reasonable component 
action still arises. One could fix the $v$ and $u$ dependence of 
$q^{2,0}$ and $q^{0,2}$ by imposing the extra constraints 
\be
\db q^{2,0} = \da q^{0,2} =0\;.  \label{extrab}
\ee
However, from the explicit structure of $\da, \db$ it immediately follows 
that
\be 
\partial_{++} V_{--}(x) =  \partial_{--} V_{++}(x) =0 \;, 
\label{badconstr1}
\ee
where $V_{++}(x)$ and $V_{--}(x)$ enter the $\theta$ expansion of 
$q^{2,0}$ and $q^{0,2}$ as the coefficients of the monomials 
$\theta^{1,0} \bar\theta^{1,0}$ and $\theta^{0,1} \bar\theta^{0,1}$, respectively (more precisely, they are first components in the 
bi-harmonic decomposition of these coefficients). 
Thus the constraints \p{extrab} lead to the dynamical 
equations-of-motion-type conditions, and so are unacceptable. 

The following relaxation of \p{extrab} proves to provide a reasonable 
extension of the constraints \p{q1}
\be  
\db q^{2,0} - \da q^{0,2} =0 \label{extracor}
\ee
(the sign minus here is a convention, one is at liberty to 
make arbitrary independent rescalings of $q^{2,0}$, $q^{0,2}$). 
It is a simple exercise to see that the set \p{q1}, \p{extracor} 
does not entail any dynamical constraints and leaves 
(32 + 32) components in $q^{2,0}$, $q^{0,2}$, 16 bosonic fields 
being physical and the remaining 16 auxiliary. One of the physical 
fields is presented, like in the case of \p{tensor}, by the 
conserved vector 
\be 
\partial_{++} V_{--}(x) - \partial_{--} V_{++}(x) =0 \; 
\Rightarrow V_{\pm}(x) = i\partial_{\pm} q (x) \;.\label{goodconstr1}
\ee
The remaining 15 bosonic fields are collected in the $\theta$ independent 
parts of $q^{2,0}$, $q^{0,2}$
\bea
q^{2,0} = q^{(ik)}(x) u^1_iu^1_k + q^{(ik)(ab)} (x) 
u^1_iu^1_k v^1_au^{-1}_b 
+ ...  \nonumber \\
q^{0,2} = q^{(ab)}(x) v^1_av^1_b + q^{(ik)(ab)} (x) 
u^1_iu^{-1}_k v^1_av^1_b
+... \;. \label{compqq}
\eea
Substituting these $q^{2,0}$, $q^{0,2}$ (with all the components 
included) into \p{q20act}, taking into account \p{goodconstr1} 
and eliminating auxiliary fields, one is left with the standard 
free $(4,4)$ supersymmetric 
$2D$ action for 16 free bosonic fields and $(16 + 16)$ fermionic 
fields of both light-cone chiralities. It is straightforward to get 
this action, so we do not quote it here. 

The most general action of the superfields $q^{2,0}, q^{0,2}$, 
by analogy with \p{Laction}, can be taken in the form 
\be
S = \int \mu^{-2,-2} L^{2,2}(q^{2,0}, q^{0,2},u, v)\;, 
\; \frac{\partial^2 L^{2,2}}{\partial q^{2,0} \partial q^{0,2}}
|_{q^{2,0} = q^{0,2}=0} \neq 0 \;. \label{q202action}
\ee
Note that we would also include into the lagrangian arbitrary powers 
of one independent harmonic derivative, say $D^{2,0}q^{0,2}$. However, repeatedly applying the constraints \p{q1}, \p{extracor} (one of their 
consequences is the vanishing of all higher-order harmonic derivatives 
of $q^{2,0}, q^{0,2}$ starting with the second-order ones) and 
integrating by parts, it is easy to show that all such terms are 
reduced to  powers of $q^{2,0}$, $q^{0,2}$. Thus the action 
\p{q202action} is indeed most general. An extension to the case 
of several copies of the pair $q^{2,0}, q^{0,2}$ goes straightforwardly.  

Along with similarities between the set of $SU(2)\times SU(2)$ harmonic constraints \p{q1}, \p{extracor} and the $SU(2)$ harmonic 
constraint \p{tensor},
there are clear differences between them. Firstly, \p{tensor} 
leaves in $L^{(++)}$ 4 physical bosonic fields while 
\p{q1}, \p{extracor} leave in $q^{2,0}, q^{0,2}$ the set of 16 ones. 
This means that the action \p{q202action} actually  propagates 
4 on-shell scalar $(4,4)$ multiplets, in contradistinction to the 
action \p{Laction} which propagates only one multiplet. Below we will 
see that this reducibility extends off shell. 

Another difference is that the constraint \p{extracor} can be explicitly 
solved in terms of scalar analytic superfield $q(\zeta, u,v)$ 
\be 
q^{2,0} = D^{2,0} q\;, \;\; q^{0,2} = D^{0,2} q \;, \label{solutcor}
\ee 
thus generalizing the solution \p{goodconstr1} to the full superfield 
level ($q(\zeta, u,v) = q(x) + ...$). Note that \p{tensor} can be 
solved only through some non-analytic prepotential \cite{tens}. 
After having been partially solved in this way, the set 
\p{q1}, \p{extracor} is reduced to 
\bea 
(D^{2,0})^2 q &=& (D^{0,2})^2 q = 0  \;\;\Rightarrow \label{constrq1}\\
q(\zeta, u, v) &=& q(x) + q^{(ik)}(x) u^1_i u^{-1}_k + 
q^{(ab)}(x) v^1_av^{-1}_b \nonumber \\ 
&& + q^{(ik)(ab)}(x)u^1_iu^{-1}_k v^1_av^{-1}_b + ...\;, 
\label{contq}
\eea
where we have written down the $\theta$ independent part of $q$ which now collects all 16 physical bosonic fields. We see that the action 
\p{q202action} is a particular representative of the general $q$ action 
\be
S_q = \int \mu^{-2,-2} L^{2,2} (q, D^{2,0}q, D^{0,2}q, u, v) 
= \int \mu^{-2,-2} \left( - D^{2,0}q D^{0,2}q + ...\right) \;.
\label{qaction}
\ee
Here we singled out the free part (the sign minus is needed  
to have the standard form of kinetic terms for physical fields) and 
took into account that the possible terms with $D^{2,0}D^{0,2}q$ can be 
reduced to those present in \p{qaction} after integrating by parts and 
exploiting the constraints \p{constrq1}. The action \p{q202action} 
corresponds to neglecting the explicit dependence on $q$ in \p{qaction} 
and leaving only harmonic derivatives of $q$. This means that 
\p{q202action} is invariant under 
arbitrary constant shifts of $q$ that is a clear symmetry of the 
constraints \p{constrq1} as well \footnote{These constraints are 
invariant under 
more general shift $q \rightarrow q + \alpha_1 + 
\alpha_2^{(ik)}u_i^1u^{-1}_k +\alpha_3^{(ab)} v^1_av^{-1}_b$ .}. The corresponding bosonic metric always has one translational $U(1)$ 
isometry, while this is not the case for the general $q$ action  
\p{qaction}.

Let us now demonstrate that the action \p{qaction} and the constraints \p{constrq1} are actually another form of the general $q^{1,1}$ action 
\p{q11action} and the constraints \p{constrq} for the case of 
4 independent $q^{1,1}$ superfields $q^{1,1\;\alpha \dot \alpha}$, 
$(\alpha, \dot \alpha = 1,2)$ (we have split the extra vector 
$SO(4)$ index into the pair of 
the doublet $SU(2)\times SU(2)$ ones). To avoid a confusion, let us 
point out that this extra $SO(4)$ commutes with $(4,4)$ supersymmetry and 
so has nothing to do with the automorphism 
$SO(4)$'s. Rather, it is an analog of the Pauli-G\"ursey $SU(2)$ 
known in the $SU(2)$ harmonic superspace formalism. 

As the harmonics $u$ and $v$ satisfy the completeness 
conditions, we can decompose $q^{1,1\;\alpha \dot \alpha}$ 
over these complete sets. We get
\be 
q^{1,1\;\alpha \dot\alpha} = q u^{1\;\alpha} v^{1\;\dot \alpha} -
q^{2,0}u^{-1\;\alpha} v^{1\;\dot \alpha} - q^{0,2}u^{1\;\alpha} 
v^{-1\;\dot \alpha} + q^{2,2} u^{-1\;\alpha} v^{-1\;\dot \alpha} 
\label{dir} 
\ee
\be
q = q^{1,1\;\alpha \dot\alpha}u^{-1}_{\alpha} v^{-1}_{\dot \alpha}\;,\; 
q^{2,0} = q^{1,1\;\alpha \dot\alpha}u^{1}_{\alpha} 
v^{-1}_{\dot \alpha}\;,\; 
q^{0,2} = q^{1,1\;\alpha \dot\alpha}u^{-1}_{\alpha} 
v^{1}_{\dot \alpha}\;,\; 
q^{2,2} = q^{1,1\;\alpha \dot\alpha}u^{1}_{\alpha} 
v^{1}_{\dot \alpha}\;, \label{invers}
\ee
where, anticipating the result, we denoted some harmonic projections of 
$q^{1,1\;\alpha \dot\alpha}$ by the same letters as the superfields 
introduced earlier. The $q^{1,1}$ constraints \p{constrq} in this 
new basis can be equivalently rewritten as the following systems 
\bea
(a)\; D^{2,0}q = q^{2,0},\;\;(b) \;D^{2,0}q^{2,0} = 0 , \;\; 
(c) \; D^{2,0}q^{0,2} = q^{2,2}, \;\; (d) \;D^{2,0}q^{2,2} = 0  
\label{1} \\
(a) \;D^{0,2}q = q^{0,2}, \;\;(b)\; D^{0,2}q^{0,2} = 0 , \;\; 
(c)\; D^{0,2}q^{2,0} = q^{2,2}, \;\; (d)\; D^{0,2}q^{2,2} = 0\;. 
\label{2} 
\eea
One sees that eqs. {\it (a)} and {\it (c)} in both systems are 
algebraic and serve to 
express the projections $q^{2,0}$, $q^{0,2}$ and $q^{2,2}$ in terms of 
the harmonic derivatives of $q$ 
\be
q^{2,0} = D^{2,0}q, \;\; q^{0,2} = D^{0,2}q, \;\; 
q^{2,2} = D^{2,0}q^{0,2} = 
D^{0,2}q^{2,0} = D^{2,0}D^{0,2} q\;\;. \label{expr}
\ee
Then eqs. {\it (b)} become just the constraints \p{constrq1}, while 
eqs. {\it (d)}  
are satisfied as a consequence both of the latter and the 
expression for 
$q^{2,2}$ in \p{expr}. So they do not imply any new restriction 
for the remaining superfield $q$. After substituting the 
expressions \p{expr} into the general action \p{q11action} 
for $q^{1,1\;\alpha \dot \alpha}$ 
in the basis \p{dir}, \p{invers}, we recover the general $q$ action 
\p{qaction}. Note that the free part of the $q^{1,1}$ 
superfield Lagrangian 
\be 
L^{free}_{q^{1,1}} \propto  q^{1,1\;\alpha \dot \alpha} 
 q^{1,1}_{\alpha \dot \alpha} = 2\left( q q^{2,2} - q^{2,0}q^{0,2} 
\right) \label{freeq}
\ee 
after integrating by parts is reduced, up to a numerical coefficient, 
to 
\be
- D^{2,0}q D^{0,2} q \;,
\ee
as should be. 

Thus we have shown that the model associated with the $SU(2)\times SU(2)$ 
analytic superfield $q$ subjected to the constraints \p{constrq1} is  
a disguised form of the theory of four self-interacting twisted 
hypermultiplets. The system of superfields $q^{2,0}$ and $q^{0,2}$ 
subjected 
to the constraints \p{q1}, \p{extracor} and described by the action \p{q202action} corresponds to a particular class of such 
self-interactions, 
with the lagrangian in \p{q11action} bearing no dependence on 
$q = q^{1,1 \;\alpha \dot\alpha} u^{-1}_\alpha v^{-1}_{\dot\alpha}$. 
This property in terms of the original field variables  $q^{1,1 \;\alpha \dot\alpha}$  can be expressed as the condition 
\be
u^{1\;\alpha} v^{1\;\dot\alpha} \frac{\partial L^{2,2}
(q^{1,1\;\beta\dot\beta}, u,v)}{\partial  q^{1,1 \;\alpha \dot\alpha}} 
= 0\;, \label{fdiffcond}
\ee  
which means the invariance of the given class of actions under the shift\footnote{Generally speaking, the invariance under \p{isometrysh} 
implies putting a full harmonic derivative 
$D^{2,0}\Lambda^{0,2} + D^{0,2}\Lambda^{2,0}$ in the r.h.s. of 
\p{fdiffcond}, where the functions $\Lambda$ {\it \'a priori} can 
bear an arbitrary dependence on $q^{1,1}, u, v$. However, it is 
easy to show that, up to full harmonic derivatives, the general 
solution to such a modified condition is  
$$L^{2,2}= \tilde{L}^{2,2} + u^{-1}_{\alpha} v^{-1}_{\dot\alpha} 
q^{1,1\;\alpha\dot\alpha} \left( D^{2,0} \tilde{\Lambda}^{0,2} 
+  D^{0,2} \tilde{\Lambda}^{2,0} \right)$$ 
where the quantities with $\;\tilde{}\;$ satisfy the condition 
\p{fdiffcond} on their own. Then, integrating by parts, one 
brings $L^{2,2}$ into the form 
$$ 
L^{2,2} = 
\tilde{L}^{2,2} - u^{1}_{\alpha} v^{-1}_{\dot\alpha}
q^{1,1\;\alpha\dot\alpha}  \tilde{\Lambda}^{0,2} -  u^{-1}_{\alpha} v^{1}_{\dot\alpha}q^{1,1\;\alpha\dot\alpha}\tilde{\Lambda}^{2,0} 
$$ 
in which it satisfies \p{fdiffcond}. Thus, without loss of generality, 
one can choose as the invariance condition just eq. \p{fdiffcond}.}  
\be 
q^{1,1\;\alpha\dot\alpha} \Rightarrow q^{1,1\;\alpha\dot\alpha} + 
\alpha_1 u^{1\;\alpha}v^{1\;\dot \alpha}\;. \label{isometrysh}
\ee
Having at our disposal the general formulas for the bosonic target 
metric and torsion in the case of general $q^{1,1}$ action 
\p{q11action} \cite{IS}, it is of course a matter of direct 
calculation to obtain them for the given particular case. 
We do not present them here.  

Our next subject will be searching for a reasonable 
$SU(2)\times SU(2)$ HSS generalization of the 
nonlinear multiplet constraint \p{harmconstrN}. Once again, requiring 
the constraints not to lead to the equations of motion fixes their 
form up to several undetermined constants 
\bea
&& D^{2,0}Q^{2,0} + \beta_1 Q^{2,0}Q^{2,0} = 0\;, 
\;\; D^{0,2}Q^{0,2} + \beta_2 Q^{0,2} Q^{0,2} = 0 \;, 
\label{nonl2020} \\
&& D^{2,0}Q^{0,2} - \beta_3 D^{0,2}Q^{2,0} + \beta_4 Q^{2,0}Q^{0,2} 
= 0 \;, 
\label{nonl2002}
\eea
where we use the capital $Q$ for the involved superfields in order to distinguish this case from the previous one. Further we separately 
consider the option when at least one of 
two free parameters in \p{nonl2020} equals zero, and the option when 
they both are non-vanishing. Heavily exploiting the commutativity 
condition \p{commut}, one can show that in the first case the 
only possibility is the set of linear constraints \p{q1}, 
\p{extracor}. In the second case, up to non-zero rescalings of 
the involved superfields, the set \p{nonl2020}, \p{nonl2002} 
can be cast into the form  
\bea
&& D^{2,0}Q^{2,0} + Q^{2,0}Q^{2,0} = 0\;, 
\;\; D^{0,2}Q^{0,2} + Q^{0,2} Q^{0,2} = 0 \label{nonlq2020} \\
&& D^{2,0}Q^{0,2} - D^{0,2}Q^{2,0} = 0 
\label{nonlq2002}
\eea
(notice the surprizing fact that the nonlinear term in 
\p{nonl2002} proves to be non-compatible with the 
self-consistency condition \p{commut}). 

What concerns the most general form of the invariant action, the 
constraints \p{nonlq2020}, \p{nonlq2002} turn out to be more 
restrictive than the linear ones \p{q1}, \p{extracor}, though not so 
severe as their $SU(2)$ HSS prototype \p{harmconstrN}. 
Though the superfield lagrangian could involve the harmonic derivative 
$D^{2,0}Q^{0,2}$ (or $D^{0,2}Q^{2,0}$), it is easy to prove that, like in 
the previous case of linear constraints, all the derivatives 
can be removed from the action by integrating by parts and 
repeatedly exploiting 
\p{nonlq2020}, \p{nonlq2002}. Then, inspecting the structure of 
the lagrangian 
as a function of $Q^{2,0}$, $Q^{0,2}$ and explicit harmonics and, 
once again, making use of the defining constraints, one can show that 
the most general superfield lagrangian in the present case is 
reducible to the form 
\be
L^{2,2}_{nonl}(Q^{2,0},Q^{0,2}, u,v) = 
Q^{2,0}\;C^{0,2}(u,v) + \sum^\infty_{n=1} 
(Q^{2,0} Q^{0,2})^n \;
C^{-2(n-1), -2(n-1)}(u,v) \;. \label{nonllagr}
\ee 
The coefficient functions in \p{nonllagr} can involve an arbitrary 
dependence on harmonics (it should be of course compatible with 
their harmonic $U(1)$ charges). Note that the first term in \p{nonllagr} 
is non-vanishing only provided the coefficient $C^{0,2}$ reveals 
a non-trivial dependence on both sets of harmonic variables.

Let us return to examining the constraints \p{nonlq2020}, 
\p{nonlq2002}. The last of them is linear, therefore in the 
present case one bosonic 
physical field is supplied by the vector $V_{\pm \pm}$ still 
subjected to the {\it linear} 
constraint \p{goodconstr1}. So the main characteristic 
feature of the $SU(2)$ nonlinear multiplet, the nonlinear 
constraint \p{nonlV}, does not 
generalize to the $SU(2)\times SU(2)$ case. Then one may 
suspect that the nonlinearity in 
the first two superfield constraints \p{nonlq2020} is also fake. 
This is indeed so, and now we wish to show that there exists 
a change of the superfield variables which brings 
\p{nonlq2020}, \p{nonlq2002}
into the linear form  \p{q1}, \p{extracor}.

This can be done in two equivalent ways. On can, e.g., firstly 
solve the 
constraint \p{nonlq2002} similarly to the previously discussed 
linear case 
\be 
Q^{2,0} = D^{2,0} \hat{Q}, \;\; Q^{0,2} = D^{0,2} \hat{Q} \;, 
\label{sol2}
\ee
and, redefining $\hat{Q}$ as 
\be 
\hat{Q} = \mbox{ln} (1 + \tilde{q}) \;, \label{redef}
\ee
reduce the remaining constraints to the form of eqs. \p{constrq1}
\be 
(D^{2,0})^2 \tilde{q} = (D^{0,2})^2 \tilde{q} = 0\;. \label{constrq2}
\ee
After this one could proceed like in the discussion of the meaning of 
\p{constrq1}: embed  \p{constrq2} into the linear set of constraints 
for four twisted superfields $\tilde{q}^{1,1\;\alpha\dot\alpha}$ 
and thus  
demonstrate that in the given case we again deal with a particular 
class of their self-interactions. 

Another way is to embed \p{nonlq2020}, \p{nonlq2002}, before 
solving them, into some extended set of nonlinear constraints with 
the superfield content $Q$, $Q^{2,0}$, $Q^{0,2}$, $Q^{2,2}$ 
characteristic 
of the projected form of some $Q^{1,1\;\alpha \dot\alpha}$. 
Consistency with the commutativity condition 
\p{commut} dictates the following unique form of such an extension 
(up to unessential constant rescalings)  
\bea 
&&(a) \;D^{2,0}Q - Q^{2,0}(1 - Q) = 0 \;, \;\; (b) \; D^{2,0}Q^{2,0} + Q^{2,0}Q^{2,0} = 0 \;, \nonumber \\ 
&&(c) \; D^{2,0}Q^{0,2} - Q^{2,2} (1-Q) + Q^{2,0}Q^{0,2} = 0 
\;, \; \;(d) \; D^{2,0} Q^{2,2} = 0 \;,\label{11} 
\eea 
\bea 
&&(a) \;D^{0,2}Q - Q^{0,2}(1 - Q) = 0 \;, \;\; (b) \; D^{0,2}Q^{0,2} + Q^{0,2}Q^{0,2} = 0 \;, \nonumber \\ 
&&(c) \; D^{0,2}Q^{2,0} - Q^{2,2} (1-Q) + Q^{2,0}Q^{0,2} = 0 \;, 
\; \;(d) \; D^{0,2} Q^{2,2} = 0 \;.\label{22}
\eea 
Like in the system \p{1}, \p{2}, some of these equations, namely, 
{\it (a)} and {\it (c)} in both sets, are algebraic and serve 
to express $Q^{2,0}$, $Q^{0,2}$, $Q^{2,2}$ in terms of $Q$. Also, 
the constraints for 
$Q^{2,2}$ are a consequence of the remainder. The constraint 
\p{nonlq2002} 
is satisfied automatically as a consequence of eqs. {\it (c)}. 
All this is visualized by passing to the new superfields (as 
before we assume that this change of superfield variables 
is invertible) 
\bea
q = \frac{Q}{1 - Q}\;, \; q^{2,0} = \frac{Q^{2,0}}{1-Q}\;, \; 
q^{0,2} = \frac{Q^{0,2}}{1-Q}\;, \; q^{2,2} = Q^{2,2}\;. \label{Qqrel}
\eea
In terms of them eqs. \p{11}, \p{22} become precisely the 
linear constraints \p{1}, \p{2}. So, $q$, $q^{2,0}$, $q^{0,2}$ 
and $q^{2,2}$ can be unified according to formulas \p{dir}, \p{invers} 
into the twisted superfield $q^{1,1\;\alpha \dot \alpha}$ with the linear 
constraint \p{constrq}. It is easy to see that $\tilde{q}$ appearing in 
\p{redef} coincides with $q$ 
\be
\tilde{q} = q, \;\;\; \hat{Q} = \mbox{ln} (1+q)\;. \label{redef1}
\ee
Note that the set of equations \p{11}, \p{22} can be rewritten as a 
nonlinear version of the constraints \p{constrq} for four twisted 
superfields $Q^{1,1 \alpha \dot \alpha}$ composed of  
$Q$, $Q^{2,0}$, $Q^{0,2}$, $Q^{2,2}$ according to eqs. \p{dir}, 
\p{invers}. 
The possibility to bring these nonlinear constraints into 
the linear form \p{constrq} 
by passing to $q^{1,1\alpha \dot\alpha}$ which is composed in the 
same way from the $q$ superfields \p{Qqrel}, reflects the fact 
that any nonlinear modification of \p{constrq} is reducible 
to the original linear form by means of some redefinition of the 
superfields $q^{1,1}$ \cite{EI}. The proof 
is based on the consistency conditions following from \p{commut}.    

It is instructive to see how the lagrangian \p{nonllagr} looks in  
new variables 
\be
L^{2,2}_{nonl}= 
\frac{q^{2,0}}{1+q}\;C^{0,2} + \sum^\infty_{n=1} 
(1+q)^{-2n}(q^{2,0} q^{0,2})^n \;
C^{-2(n-1), -2(n-1)} \;. \label{nonllagr1}
\ee 
Among all possible actions of $q$, $q^{2,0}$, $q^{0,2}$ it is 
distinguished in having the scaling isometry 
\be 
\delta q = \alpha (1+q), \;\delta q^{2,0} = \alpha q^{2,0}, \; 
\delta q^{0,2} = \alpha q^{0,2}\;, \;\delta q^{2,2} = \alpha q^{2,2}\;,  
\label{isom2}
\ee
which, in the original variables, affects only $Q$ and $Q^{2,2}$ 
\be
\delta Q = \alpha (1-Q), \;\delta Q^{2,2} = \alpha Q^{2,2} \;, 
\label{isom2q}
\ee
leaving $Q^{2,0}$, $Q^{0,2}$ intact. It is easy to check the 
covariance of \p{11}, \p{22} or \p{1}, \p{2} under 
these transformations. Note that this isometry is 
different from the pure shifting one which is inherent to the action \p{q202action}. 
In the language of the superfield $q^{1,1\;\alpha \dot\alpha}$ it is 
represented as 
\be
\delta q^{1,1\;\alpha\dot \alpha} = \alpha \left( 
q^{1,1\;\alpha\dot \alpha} + 
u^{1\;\alpha} v^{1 \;\alpha} \right) \;, \label{isomq11}
\ee           
that is to be compared with eq. \p{isometrysh}. On the superfield 
$\hat{Q} = \mbox{ln}(1+q)$ it is realized by pure translations, however $\hat{Q}$ satisfies nonlinear constraints. On the other hand, the set of 
linear constraints \p{1}, \p{2} reveals invariance under shifts 
$$
q \rightarrow q + \alpha', \;\; \;\alpha' = const\;, 
$$
however the original nonlinear set of constraints 
\p{nonlq2020}, \p{nonlq2002} is not closed under such transformations ($Q^{2,0}$, $Q^{0,2}$ transform 
through a superfield $Q$ which is not explicitly present in 
\p{nonlq2020}, \p{nonlq2002}). Correspondingly, the 
lagrangians \p{nonllagr}, \p{nonllagr1} do not respect this 
second isometry. In other words, the 
linear and nonlinear $SU(2)\times SU(2)$ multiplets are adapted for 
describing {\it different} subclasses of general self-interactions 
of $q^{1,1\alpha\dot\alpha}$.   
  
Note that the first term 
in \p{nonllagr1}, after substituting $q^{2,0} = D^{2,0}q$ 
and integrating by parts, is reduced to 
$$
-\mbox{ln} (1+q)\; D^{2,0}C^{0,2}.
$$   
It produces a non-standard kinetic term for $q$ 
$$
\sim \left( D^{2,0}D^{0,2}Q D^{2,0}D^{0,2}q \right) 
C^{-2,-2}|;,\;\; C^{0,2} \equiv 
D^{2,0}(D^{0,2})^2 C^{-2,-2}\;.
$$
Its diagonalization implies a complicated redefinition of 
physical fields 
by the components of the $SU(2)$ breaking tensor 
$C^{-2,-2} = C^{(ik)(ab)} u^{-1}_i u^{-1}_k v^{-1}_a v^{-1}_b$.  
On the other hand, the second sequence of self-interactions 
in \p{nonllagr1} contains the standard kinetic term. It comes 
from the first term in the sum 
\be
L^{2,2 (1)}_{nonl} = (1+q)^{-2} q^{2,0}q^{0,2} C^{0,0} = 
- q^{2,0}q^{0,2} 
+ ...\;.
\label{conf}
\ee      

\setcounter{equation}{0}

\section{Discussion}

Despite the fact that the constrained $SU(2)\times SU(2)$ 
harmonic analytic 
superfields defined above are basically equivalent to 
four twisted $(4,4)$ superfields, the use of these off-shell 
representations has some merits which we wish to briefly 
outline here. 

One of these advantages is related to the use of $SU(2)\times SU(2)$ 
harmonic analog of nonlinear multiplet. It turns out that $2D$, $N=4$ superconformal group admits a simple realization in terms of this 
superfield, and it becomes easy to construct the corresponding 
invariant action. 

As was discussed in \cite{IS}, 
in the $SU(2)\times SU(2)$ analytic HSS one can realize 
two different ``small'' $N=4$, $SU(2)$ superconformal groups 
(in each light-cone sector), 
having as their closure the ``large'' $N=4$, $SO(4)\times U(1)$ 
superconformal group. One of these $N=4$, $SU(2)$ groups does not affect 
harmonic variables, the superfields $q^{1,1}$ and any their harmonic 
projections behave as scalars with respect to it. The analytic 
superspace integration measure is also invariant. So all the actions 
considered above trivially enjoy invariance under this superconformal 
group. Another 
$N=4$, $SU(2)$ superconformal group affects the harmonic variables 
\cite{{DS},{IS}}
\be 
\delta u^{1}_i = \Lambda^{2,0}u^{-1}_i\;, \; \delta u^{-1}_i = 0\;;
\;\;\;\; 
\delta v^{1}_a = \Lambda^{0,2}v^{-1}_a\;, \; \delta v^{-1}_a = 0 \;,\label{uvtransf}
\ee
\bea
\delta D^{2,0} &=& - \Lambda^{2,0} D^0_u\;, \;\; 
\delta D^{0,2} = -\Lambda^{0,2} D^0_v\;, \label{Dtransf} \\ 
D^{2,0}\Lambda^{2,0} &=& D^{0,2}\Lambda^{2,0} = D^{0,2}\Lambda^{0,2} = D^{2,0}\Lambda^{0,2} = 0\;, \label{lamcons}
\eea
where $D^0_u$, $D_v^0$ are the left and right harmonic $U(1)$ charge 
operators. 
It will be essential for us that 
the parameter superfunctions $\Lambda^{2,0}$ ($\Lambda^{0,2}$) 
depend only 
on the coordinates $z_{++}, \theta^{1,0\;\underline{i}}, u^{\pm}_i$ 
($z_{--}, \theta^{0,1\;\underline{a}}, v^{\pm}_a$) and satisfy 
the harmonic constraints just written. The realization of this group 
on other analytic superspace coordinates besides the harmonic 
ones can be found in \cite{IS}. 
The analytic superspace integration measure is invariant in this case 
as well.
Also we will need the fact that the superfield $q^{1,1}$ transfoms 
with an 
analytic weight under this group 
\bea 
\delta q^{1,1\;\alpha \dot\alpha} &=& \left( 
\Lambda_L + \Lambda_R \right) 
q^{1,1\;\alpha \dot\alpha}, \;\Lambda^{2,0} \equiv D^{2,0}\Lambda_L, 
\;\Lambda^{0,2} \equiv D^{0,2}\Lambda_R\;, \label{qconf} \\
D^{0,2}\Lambda_L &=& D^{2,0}\Lambda_R = 0 \label{lamcons1}
\eea    
(this transformation law unambiguously follows from requiring the 
covariance of the harmonic constraint \p{constrq}).

It is not so easy to construct the action of $q^{1,1}$s invariant 
under this 
second $N=4$, $SU(2)$ superconformal group. For one $q^{1,1}$, 
as shown in 
\cite{IS}, the unique invariant action is that of $N=4$ $SU(2)\times 
U(1)$ WZNW sigma model
\be \label{wznw}
S_{wznw} = {1\over \kappa^2} \int \mu^{-2,-2} \hat{q}^{1,1}\hat{q}^{1,1} 
\left( \frac{\mbox{ln} (1+X)}{X^2} - \frac{1}{(1+X)X}\right)\;, 
\ee  
where 
$$
\hat{q}^{1,1} \equiv q^{1,1} - c^{1,1}, \; X \equiv c^{-1,-1}
\hat{q}^{1,1}, \;
c^{\pm 1,\pm 1} = c^{ia}u^{\pm1}_iv^{\pm1}_a\;, c^{ia} = const, \;
c^{ia}c_{ia} = 2\;. 
$$
Despite the presence of an extra quartet constant $c^{ia}$ in the analytic 
superfield lagrangian, the action \p{wznw} does not depend on $c^{ia}$, 
as it is invariant under arbitrary rescalings and $SU(2)\times 
SU(2)$ rotations of this constant. The invariance of \p{wznw} under 
the second $N=4$ superconformal group which is realized on 
$\hat{q}^{1,1}$ as 
\be 
\delta \hat{q}^{1,1} = (\Lambda_L + \Lambda_R)(\hat{q}^{1,1} + c^{1,1}) - 
\Lambda^{2,0}c^{-1,1} - \Lambda^{0,2}c^{1,-1} \label{transfhat}
\ee
is not manifest. It is a tedious though straightforward exercise to check 
this invariance.

It turns out that for $q^{1,1\;\alpha\dot\alpha}$ one can 
construct the action almost {\it manifestly} invariant 
under this superconformal group. This action belongs to 
the subclass of actions \p{nonllagr}, \p{nonllagr1} 
associated with the  ``$SU(2)\times 
SU(2)$ nonlinear multiplet'' $Q^{2,0}, \;Q^{0,2}$ defined by the 
constraints 
\p{nonlq2020}, \p{nonlq2002}. Using the transformation rules \p{Dtransf} 
it is easy to check that these constraints are 
consistent with the following transformation properties of the involved 
superfields 
\be
\delta Q^{2,0} = \Lambda^{2,0}\;, \;\; 
\delta Q^{0,2} = \Lambda^{0,2}\;. 
\label{q20conf}
\ee
Then the particular representative of the lagrangians \p{nonllagr}, 
\be 
L^{2,2(1)}_{conf} = Q^{2,0} Q^{0,2}\; C = 
\frac{q^{2,0}q^{0,2}}{(1+q)^2}\; C = 
D^{2,0}\;\mbox{ln} (1+q)\; D^{0,2}\;\mbox{ln} (1+q)\;C\;, 
\; C = const \;, 
\label{lagconf} 
\ee
is obviously shifted by a full harmonic derivative under \p{q20conf} 
as a consequence of the structure of $\Lambda^{2,0}$, 
$\Lambda^{0,2}$ \p{qconf} and the harmonic constraints 
\p{lamcons}, \p{nonlq2002}. Ascribing to the 
superfields defined by eqs. \p{Qqrel} the following superconformal transformation properties
\bea
\delta q^{2,0} &=& (\Lambda_L + \Lambda_R) q^{2,0} +  
\Lambda^{2,0} (1 + q), \;\delta q^{0,2} = (\Lambda_L + 
\Lambda_R) q^{0,2} + 
\Lambda^{0,2} (1+q), \nonumber \\
\delta q &=& (\Lambda_L + \Lambda_R) (1+q), \; 
\delta q^{2,2} = (\Lambda_L + \Lambda_R) q^{2,2} + 
\Lambda^{2,0} q^{0,2} 
+\Lambda^{0,2}q^{2,0}\;, \label{q20conf1}
\eea
we see that they are consistent with \p{q20conf} and the standard 
transformation properties of $q^{1,1\alpha \dot\alpha}$ 
\be
\delta q^{1,1\alpha\dot\alpha} = (\Lambda_L + \Lambda_R) 
q^{1,1\alpha\dot\alpha}\;,  \label{Qconf}  
\ee
provided the following identification has been made
\be
q^{1,1\alpha \dot\alpha} u^{-1}_\alpha v^{-1}_{\dot \alpha} 
\equiv 1 + q, \; 
q^{1,1\alpha \dot\alpha} u^{1}_\alpha v^{-1}_{\dot \alpha} 
\equiv q^{2,0}, \;  
q^{1,1\alpha \dot\alpha} u^{-1}_\alpha v^{1}_{\dot \alpha}  
\equiv q^{0,2}, \;
q^{1,1\alpha \dot\alpha} u^{1}_\alpha v^{1}_{\dot \alpha}  
\equiv q^{2,2}.
\label{canQ} 
\ee
The transformation properties of the additional superfields 
$Q$ and $Q^{2,2}$ 
entering the extended set of nonlinear constraints \p{11}, \p{22} 
can be deduced directly from the relations \p{Qqrel}. It is 
straightforward 
to check the covariance of this set (as well as of its linear 
counterpart 
for the superfeilds $q, q^{2,0},...$) under the superconformal 
transformations. It would be interesting to examine in detail the 
component content of the action \p{lagconf}. Note that the above 
transformation properties 
can be easily extended to the full conformal $(4,4)$ supergravity in 
$SU(2)\times SU(2)$ harmonic superspace \cite{BI}. 
Then the multiplet $Q^{2,0}$, $Q^{0,2}$ can serve as a compensator 
reducing 
this supergravity to a kind of off-shell Einstein $(4,4)$ one. 

Our next comment concerns the dual formulations of the superfield 
actions 
presented. As was shown in \cite{IS}, the twisted multiplet 
constraints  
\p{constrq} can be implemented in the action with the help of 
appropriate 
analytic superfield lagrange multipliers to yield a new off-shell 
representation of the $q^{1,1}$ action in terms of unconstrained 
analytic 
superfields with an infinite number of auxiliary fields. For the case 
of $q^{1,1\;\alpha \dot\alpha}$, the action with the lagrange 
multipliers 
terms added  reads 
\be 
L^{2,2}_{\omega, q} = \omega^{-1,1\;\alpha \dot\alpha} D^{2,0} q^{1,1}_{\alpha\dot\alpha} + \omega^{1,-1\;\alpha \dot\alpha} 
D^{0,2} q^{1,1}_{\alpha \dot\alpha} + L^{2,2}(q^{1,1},u,v)\;. 
\label{lagrmul1}
\ee    
Eliminating $q^{1,1\;\alpha\dot\alpha}$ by their equations of motion 
at expense of $\omega$ superfields, one expresses the action 
in terms of 
the latter and so gets another off-shell representation of this 
action.  
The basic feature of this type of duality transformation is the 
gauge invariance 
\be
\delta \omega^{1,-1\;\alpha\dot\alpha} = D^{2,0} 
\sigma^{-1,-1\;\alpha \dot\alpha}\;, \; 
\delta \omega^{-1,1\;\alpha\dot\alpha} = - D^{0,2} 
\sigma^{-1,-1\;\alpha \dot\alpha},  
\label{gauge1} 
\ee
with $\sigma^{-1,-1\;\alpha \dot\alpha}$ being unconstrained analytic 
superfield parameters. It serves to reduce the set of 
physical fields in the $\omega$ superfields just to the on-shell 
content of twisted multiplet, thus ensuring the on-shell equivalence 
of the original and dual formulations of the latter.

This kind of duality is crucially different from the duality associated 
with tensor multiplet discussed in the context of $SU(2)$ HSS in Sect. 2. 
Due to the presence of constrained $2D$ vector
in the superfield $L^{(++)}$, one of the bosonic on-shell 
degrees of freedom in this case is represented in essentailly different 
ways in the initial and dual formulations. As the result, the 
relevant duality transformation  not only changes the off-shell content 
of the theory, but also affects the on-shell structure of the action, 
yielding sigma model with a different target space geometry, in 
accord with the general concept of abelian T-duality in $2D$ 
sigma models \cite{{Buscher},{Tdual}}. The presence of 
abelian isometry in both the $L^{(++)}$ action and its dual is 
most essential for the related duality to fall in this general class.  

On the contrary, the duality associated with the $q^{1,1}$ action is 
not ``genuine'' in the sense that it merely changes the off-shell 
structure of the action. After fixing an appropraite gauge 
with respect to \p{gauge1} and eliminating an infinite tail 
of auxiliary fields, the $\omega$ action gives rise to the precisely 
same component on-shell sigma model action as the original constrained 
$q^{1,1}$ action. Such a duality exists irrespective of whether the 
$q^{1,1}$ action possesses any isometry. The obvious reason 
why the standard T-duality mashinery does not apply to this case 
is the absence of constrained vectors in the superfield $q^{1,1}$ 
subjected to constraints \p{constrq}.

The superfield systems which we discussed in Sect. 3 correspond to 
particular 
classes of $q^{1,1}$ actions with abelian translational isometries. 
This matches with the fact that the sets of constraints 
\p{q1}, \p{extracor} and \p{nonlq2020}, \p{nonlq2002} by which we 
originally defined these superfields imply the existence of 
constrained vectors among their irreducible components. Thus 
we can expect the existence of the ``genuine'' duality transformation 
in these cases, 
along with the standard $q^{1,1}$ duality related to treating    
\p{q1}, \p{extracor} or \p{nonlq2020}, \p{nonlq2002} as a subclass of 
twisted multiplet constraints \p{constrq}. In other words, 
when working 
with the harmonic projections of $q^{1,1\alpha\dot\alpha}$, we are at 
freedom either to include into the action the whole set of constraints, 
and this amounts to the standard $q^{1,1}$ duality, or to 
implement with 
lagrange multipliers only part of them, deducing the 
remainder by solving these few basic constraints. Both  
procedures preserve manifest $(4,4)$ supersymmetry, 
however lead to essentially different dual actions. 

Let us apply to the system associated with the action 
\p{q202action} and constraints \p{q1}, \p{extracor}. It is the 
simplest one because the relevant $U(1)$ isometry is realized 
in this case 
as a pure shift of the superfield $q$. We could implement the 
extended set of constraints \p{1}, \p{2} in the action with lagrange multipliers; what we 
would obtain in this case is the same lagrange multiplier term as in \p{lagrmul1} but written in terms of the bi-harmonic projections of $q^{1,1\;\alpha\dot\alpha}$. On the 
other hand, we can do the same trick with the original set of 
constraints 
\p{q1}, \p{extracor} without explicitly solving \p{extracor} 
through the superfield $q$. We get in this way the following 
superfield action 
\bea
\tilde{S} &=& \int \mu^{-2,-2} \left[ L^{2,2}
(q^{2,0},q^{0,2},u,v) + 
\omega^{-2,2}D^{2,0}q^{2,0} + \omega^{2,-2}D^{0,2}q^{0,2} 
\right. \nn \\ 
&& + \left. \omega \left( D^{2,0}q^{0,2} - D^{0,2}q^{2,0} 
\right) \right]\;.
\label{dualist}
\eea
Varying with respect to $\omega$'s gives the set \p{q1}, 
\p{extracor} which, 
after solving eq. \p{extracor}, leads to the set \p{constrq1} 
or the equivalent 
one \p{1}, \p{2}. As the result, we end up with a particular 
$U(1)$ invariant class of the $q^{1,1\alpha\dot\alpha}$ actions. 
On the 
other hand, varying with respect to $q^{2,0}$, $q^{0,2}$ 
which are now unconstrained, we get 
\be 
q^{2,0} = -D^{2,0}\omega - D^{0,2}\omega^{2,-2} + ... 
\;,\;\;
q^{0,2} = D^{0,2}\omega - D^{2,0}\omega^{-2,2} + ...\;, 
\label{qomexpr}
\ee
where dots stand for the terms of higher order in superfields. After 
substituting this back into the action \p{dualist} one obtains the 
dual of \p{q202action} in terms of 
unconstrained analytic superfields $\omega, \omega^{2,-2}, 
\omega^{-2,2}$. The number of physical fields in both formulations 
can be checked to coincide due to the invariance of the action 
\p{dualist} and expressions \p{qomexpr} under the gauge transformations 
(cf. \p{gauge1})
\be \label{gauge2}
\delta \omega = - D^{2,0}D^{0,2}\sigma^{-2,-2}\;, \; 
\delta \omega^{2,-2} = (D^{2,0})^2 \sigma^{-2,-2}\;, \;  
\delta \omega^{-2,2} = -(D^{0,2})^2 \sigma^{-2,-2} \;, 
\ee
with $\sigma^{-2,-2}$ being an unconstrained analytic superfield 
parameter. 
However, one isoscalar bosonic degree of freedom is represented 
in different 
ways in both formulations: in the original setting by the 
vector $V_{\pm\pm}$
subjected to the constraint \p{goodconstr1} which can be 
solved through a field $q(x)$, and in the dual formulation 
by the first component $\omega_0(x)$ 
of the superfield $\omega$ 
\be \label{defom}
\omega(\zeta, u, v) = \omega_0 (x) + ... \;.
\ee  
Thus we are facing the situation quite similar to the 
interplay between the original constrained and dual 
formulations of the $(4,4)$, $2D$ tensor 
multiplet action in the $SU(2)$ HSS. 

To give a feeling of the basic geometric features of this 
interplay in the 
present case, 
let us briefly describe some component results concerning the 
structure of the bosonic part of the action \p{dualist}. After 
eliminating an infinite tower of auxiliary fields coming from 
$q^{2,0}, q^{0,2}$ and the 
lagrange multipliers (with fixing an appropriate WZ gauge with 
respect to the gauge freedom \p{gauge2}), this part can be 
written entirely in terms of the 
fields $q^{(ik)}(x), q^{(ab)}(x), q^{(ik)(ab)}(x)$ defined 
in eq. \p{compqq}, 
the $2D$ vector field $V_{\pm\pm}(x)$ and the scalar 
field $\omega_0(x)$ 
\bea 
\tilde{S}_{bos} &=& \int d^2x [ \;i\omega_0\left( 
\partial_{++}V_{--} - \partial_{--}V_{++}\right) \nn \\
&& + \int [du dv]\; L_{bos} (q^{(ik)}, q^{(ab)}, q^{(ik)(ab)}, 
V_{\pm\pm}, u,v)\;] \;. \label{bos}
\eea
For the time being we do not explicitly specify the lagrangian 
$L_{bos}$. It involves harmonic integrals of the 
three functions 
\be 
A^{-2,2} \equiv \frac{\partial^2 L^{2,2}}{\partial q^{2,0} 
\partial q^{2,0}}|\;, \;A^{2,-2} \equiv 
\frac{\partial^2 L^{2,2}}{\partial q^{0,2} \partial q^{0,2}}|\;,\; 
A \equiv -\frac{\partial^2 L^{2,2}}{\partial q^{0,2} 
\partial q^{2,0}}| 
\label{funct} 
\ee
multiplied by appropriate monomilas of the harmonics $u$, $v$ 
(hereafter, the symbol $|$ means restricting to the $\theta$ 
independent parts of the superfields and, for $q^{2,0}$, $q^{0,2}$, 
also keeping only 
the physical parts \p{compqq}). Note that we can choose the WZ 
gauge so that 
\be
\omega | = \omega_0(x) + \hat \omega_1 (x,u) + \hat \omega_2 (x, v)\;, 
\ee
where the objects with ``hat'' start with monomilas 
$u^1_{(i}u^{-1}_{k)}$, 
$v^1_{(a}v^{-1}_{b)}$, respectively. Then the non-dynamical 
harmonic equations of motion
\be 
\frac{\partial L^{2,2}}{\partial q^{2,0}}| - \partial^{2,0} 
\omega^{-2,2}| 
+ \partial^{0,2} \hat \omega_2  = 0\;, \; 
\frac{\partial L^{2,2}}{\partial q^{0,2}}| - \partial^{0,2} 
\omega^{2,-2}| 
- \partial^{2,0} \hat \omega_1  = 0\;, \label{harmur}
\ee
express the $\theta$ independent parts of the lagrange multipliers 
through the fileds $q^{(ik)}(x)$, $q^{(ab)}(x)$, $q^{(ik)(ab)}(x)$ and 
simultaneously establish equivalence relations between these fields 
and the appropriate lowest isospin components in the bi-harmonic 
expansions 
of $\omega^{-2,2}|$, $\omega^{2,-2}|$ and $\omega|$: 
\bea 
\hat{\omega}_1 (x,u) &=& - q^{(ik)}(x)u^1_i u^{-1}_k + ... \;, 
\;\;\;
\hat{\omega}_2 (x,v) = q^{(ab)}(x)v^1_a v^{-1}_b + ...\;, 
\nn \\ 
\omega^{2,-2}(x,u,v) &=& -{1\over2} 
q^{(ik)(ab)}(x)u^1_iu^1_k v^{-1}_a v^{-1}_b 
+ ... \;, \nn \\ 
\omega^{-2,2}(x,u,v) &=& -{1\over2} q^{(ik)(ab)}(x)
u^{-1}_iu^{-1}_k v^{1}_a v^{1}_b + ... \;.\label{canrel}
\eea
These relations justify the choice of the same 
fields $q^{(ik)}(x), q^{(ab)}(x)$, $q^{(ik)(ab)}(x)$ to 
represent the 
appropriate physical bosonic degrees of freedom in the 
original and dual formulations. 

The bosonic on-shell action in terms of 16 fields 
$q(x)$,  $q^{(ik)}(x), q^{(ab)}(x)$, $q^{(ik)(ab)}(x)$ 
corresponding to the 
initial formulation comes out if one varies with respect to 
$\omega_0(x)$ 
in \p{bos} and substitutes back the purely gradient solution 
\p{compqq} for 
$V_{\pm\pm}$. On the other hand, varying with respect to  
$V_{\pm\pm}$ yields the dual action with the 16th scalar 
represented 
by the lagrange multiplier field  $\omega_0$. So the on-shell 
geometry in 
both formulations is not the same; the relevant geomteric 
quantities, viz. 
the components of the metric and torsion potential, are interrelated 
according to the general abelian $T$-duality 
formulas \cite{{Buscher},{Tdual}}. 

In order to precisely see how this occurs, let us consider a 
simplified situation 
when all the bosonic fields except for 
$V_{\pm\pm}(x)$, $q^{(ab)}(x)$, 
$\omega_0 (x)$ are put equal to zero
\be
q^{(ik)} = q^{(ik)(ab)} = 0 \label{red1}
\ee
and the functions defined in \p{funct} are assumed to bear 
no dependence 
on these fields.
In this reduced case the second 
term in the integrand in \p{bos} is expressed through one function 
\be 
A = A (q^{(ab)}v^1_av^1_b, u, v) = 1 +...\label{red}
\ee 
and is given by the following expression 
\bea
\int [dudv] L_{bos} &=& -{\cal A} \left( V_{++}V_{--} -{1\over 2} 
\partial_{++}q^{(ab)}\partial_{--}q_{(ab)} \right) \nn \\
&& - i{\cal A}_{(ab)}\left( V_{--}\partial_{++}q^{(ab)} - 
V_{++}\partial_{--} 
q^{(ab)} \right)\;, \label{redbos}
\eea
with 
\be
{\cal A} \equiv \int [dudv] A\;, \;\; {\cal A}_{(ab)} \equiv 
\int [dudv] A v^1_{(a}v^{-1}_{b)}\;. \label{defV}
\ee
After varying with respect to $\omega_0$ and solving the resulting 
constraint as in eq. \p{compqq}, one gets the sigma model action 
with torsion 
\bea
\tilde{S}_{bos}^{red} &=& \int d^2x \;[\; G_{00} \left(  
\partial_{++}q_0 \partial_{--}q_0 + 
{1\over 2} \partial_{++}q^{(ab)}\partial_{--}q_{(ab)} \right) \nn \\
&&+ \;  
B_{0(ab)}\left( \partial_{++}q_0 \partial_{--}q^{(ab)} -          
\partial_{--}q_0 \partial_{++}q^{(ab)} \right)\; ]\;, \label{bos1}
\eea
\be 
G_{00} = {\cal A}\;, \;\;B_{0(ab)} = -{\cal A}_{ab}\;.
\ee
Varying with respect to $V_{\pm\pm}$ and substituting the result 
into the 
action entirely eliminate the torsion term from the latter, 
and the dual form of \p{bos1} proves to be as follows 
\bea
\tilde{S}_{dual}^{red} &=& \int d^2x \;[\; ({\cal A})^{-1} 
\left( \partial_{++}\omega_0 + \partial_{++}q^{(ab)}{\cal A}_{(ab)} 
\right) 
 \left( \partial_{--}\omega_0 + \partial_{--}q^{(ab)}{\cal A}_{(ab)} 
\right) 
\nn \\ 
&&+ \;{1\over 2} {\cal A}\; \partial_{++}q^{(ab)}
\partial_{--}q_{(ab)}\;]\;.
\label{hkaction}
\eea
What we have got is the general hyper-K\"ahler $4$ dimensional 
sigma  model with one translational isometry. Indeed, in accord 
with the 
general parametrization of such metrics \cite{reviewHK}, 
the function ${\cal A}$ by construction satisfies the 
Laplace's equation 
\be \label{laplace}
\partial^{(ab)}\partial_{(ab)} {\cal A} = 0\;,\;\;\;\partial_{(ab)} 
\equiv 
\frac{\partial}{\partial q^{(ab)}}\;, 
\ee
and is none other then the twistor transform for the general 
solution of 
the latter, while ${\cal A}_{(ab)}$ is related to 
${\cal A}$ by the well-known equation 
\be 
\partial_{(ab)}{\cal A}_{(cd)}- 
\partial_{(cd)} {\cal A}_{(ab)}  = {1\over 2} 
\left( \partial_{(ca)} {\cal A}\;\epsilon_{db} + 
+ \partial_{(db)} {\cal A}\; \epsilon_{ca} \right)\;. 
\label{relAA}
\ee         
The relation between the quantities entering the torsionless 
hyper-K\"ahler sigma model action \p{hkaction} and those 
present in the action \p{bos1} 
is given by the standard $T$ duality relations. In the same 
way, another reduction 
\be
q^{(ab)} = q^{(ik)(ab)} = 0 \label{red2}
\ee
also yields a general $4$ dimensional hyper-K\"ahler 
manifold with one translational isometry. 
 
Thus we have proven that the general action of four 
twisted superfields $q^{1,1\;\alpha\dot\alpha}$ with one 
purely translational 
isometry leads in the bosonic sector to the torsionful 
sigma model, 
such that the relevant $16$ dimensional bosonic manifold 
admits two reductions 
to the $4$ dimensional hyper-K\"ahler submanifolds with 
the same isometry. 
The hyper-K\"ahler nature of these submanifolds is visualized  
by the duality transformation which can be formulated in 
a manifestly $(4,4)$ 
supersymmetric way, based upon the ``master action'' 
\p{dualist}. Note that 
in the case of one twisted supermultiplet no reduction 
to hyper-K\"ahler 
manifolds exists. Let us also point out that the above 
duality transformation certainly does not eliminate 
all the torsion terms from the original non-reduced 
sigma model action, 
it is impossible to remove the components of the 
torsion potential $B_{(ab) (ik)(cd)}$ and others. 
This demonstrates that the torsion is intrinsically 
inherent to the $(4,4)$ models we consider, 
in contrast, e.g.,  to the "fake" torsion in the 
reduced action \p{bos1}. 

Our last remark will be on another, somewhat puzzling relation 
of the multiplets $q^{2,0}, q^{0,2}$ and $Q^{2,0}, Q^{0,2}$ to a  
{\it single} twisted superfield $q^{1,1}$. 

After some algebra which involves integrating by parts 
with respect 
to harmonics and making use of the relation
$$
c^{1,1}c^{-1,-1} - c^{1,-1}c^{-1,1} = 1,
$$
one can cast the $N=4$, $SU(2)\times U(1)$  WZNW action \p{wznw} 
into the following suggestive form 
\bea
S_{wznw} &=& {1\over \kappa^{2}} \int \mu^{-2,-2} 
\left(\hat{q}^{1,1} \right)^2 
\frac{c^{-1,1}c^{1,-1}}{(1+X)^2} \nn \\ 
&=& {1\over \kappa^2} \int \mu^{-2,-2}\; 
D^{2,0}\;\mbox{ln} (1 + X)\; D^{0,2}\; \mbox{ln} (1+X) \;. 
\label{wznwnew}
\eea  
In this form the action looks literally as \p{lagconf} with 
the following identifications 
\bea
\tilde{Q}^{2,0} &=& \frac{c^{1,-1}\hat{q}^{1,1}}{(1+X)}\;, \; 
\tilde{Q}^{0,2} = \frac{c^{-1,1}\hat{q}^{1,1}}{(1+X)} \;, 
\tilde{Q} = \frac{X}{1+X} \label{strangeQ} \\
\tilde{q}^{2,0} &=& c^{1,-1}\hat{q}^{1,1}\;,\; \tilde{q}^{0,2} = c^{-1,1}\hat{q}^{1,1}\;, \; 
\tilde{q} = X\;. \label{strangeq}
\eea
It is a simple exercise to check that these objects obey, 
respectively, the sets of constraints \p{nonlq2020}, 
\p{nonlq2002} and \p{q1}, \p{extracor} as a 
consequence of the $q^{1,1}$ constraint \p{constrq} 
and the relations 
$$
D^{2,0}c^{-1,\pm 1} = c^{1,\pm 1}\;, \;D^{0,2}c^{\pm 1, -1} 
= c^{\pm 1, 1}\;.
$$
Then the relations \p{strangeQ}, \p{strangeq} are precisely 
eqs. \p{Qqrel}. 

This correspondence between $\hat{q}^{1,1}$ and the $\tilde{Q}$ and 
$\tilde{q}$ superfields can be understood as follows. 

We will specialize to the $\tilde{q}^{2,0}, \tilde{q}^{0,2} 
\leftrightarrow 
q^{1,1}$ correspondence, because the existence of an analogous 
one between $q^{1,1}$ and the $\tilde{Q}$ 
superfields automatically follows from the map \p{Qqrel}. 
Let us consider the following extension of the 
constraints \p{q1}, \p{extracor} 
\bea 
D^{2,0}q^{2,0} &=& 0\;, \; D^{0,2}q^{0,2} = 0\;, \;  
D^{2,0}q^{0,2} - D^{0,2}q^{2,0} = 0\;, \label{star} \\
c^{1,-1}q^{0,2} - c^{-1,1}q^{2,0} &=& 0\;, \label{nov} 
\eea
where the charged constants $c^{\pm 1, \mp 1}$ are of 
the same type as above. Defining 
\be
\hat{q}^{1,1} \equiv c^{-1,-1} D^{2,0}q^{0,2} - c^{-1,1} q^{2,0}\;,
\label{defq}
\ee
it is easy to check that this object satisfies 
the constraints \p{constrq}, 
$$
D^{2,0}\hat{q}^{1,1} = D^{0,2}\hat{q}^{1,1} = 0 
$$
as a consequence of \p{star}, \p{nov}. Also, one can check 
that $q^{2,0}$ and $q^{0,2}$ satisfying \p{star}, \p{nov} are 
expressed through $q^{1,1}$ \p{defq} by the relations  \p{strangeq}. 
Thus we have proven one-to-one correspondence between the 
superfields $q^{2,0}$, $q^{0,2}$ subjected to the 
constraints \p{star}, 
\p{nov} and the twisted superfield $q^{1,1}$. Quite analogously, we 
reveal one-to-one correspondence between $q^{1,1}$ and the 
superfields $Q^{2,0}$, $Q^{0,2}$ subjected to nonlinear constraints  
\p{nonlq2020}, \p{nonlq2002} and the same additional 
constraint \p{nov} 
\be
c^{1,-1}Q^{0,2} - c^{-1,1}Q^{2,0} = 0\;. \label{novQ}             
\ee
In this case $\hat{q}^{1,1}$ is defined by 
\be
\hat{q}^{1,1} = \frac{A^{1,1}}{1-c^{-1,-1}A^{1,1}}\;, \;\;\; 
A^{1,1} = c^{-1,-1}\left( D^{2,0}Q^{0,2} + Q^{2,0}Q^{0,2} \right) 
- c^{-1,1}Q^{2,0}\;. 
\ee
Thus we have found a new description of $q^{1,1}$ in terms of the 
superfields $q^{2,0}, q^{0,2}$ or $Q^{2,0}$, $Q^{0,2}$ with one 
additional algebraic constraint \p{nov}, \p{novQ}. Note that 
the latter is not covariant with respect to the simple 
realization of 
$N=4$ superconformal group given by the transformation 
laws \p{q20conf}, 
\p{q20conf1}. At the same time, the extended set of constraints \p{star}, 
\p{nov} and its $Q$ counterpart  
are covariant under a  
more complicated realization of this group which is induced on 
$Q^{2,0}$, $Q^{0,2}$ and $q^{2,0}, q^{0,2}$ by the transformation law 
\p{transfhat} of $\hat{q}^{1,1}$ through the correspondence 
\p{strangeQ}, \p{strangeq}. The interplay between these two 
relaizations of $N=4$ superconformal group remains to be 
understood.

\setcounter{equation}{0}
\section{Conclusion}
In this paper we have constructed the $SU(2)\times SU(2)$ 
HSS analogs of the 
standard tensor and nonlinear off-shell $(4,4)$ multiplets. 
These new 
$(4,4)$ multiplets are represented by the properly constrained 
$SU(2) \times SU(2)$ analytic harmonic superfields 
$q^{2,0}, q^{0,2}$ and $Q^{2,0}, Q^{0,2}$,  comprise 
$(32+32)$ component fields (one of their 16 
physical bosonic fields is supplied by a constrained vector) 
and yield 
$(4,4)$ sigma models with torsion. The relevant actions 
can be equivalently 
given in terms of four twisted $(4,4)$ superfields 
$q^{1,1\;\alpha\dot\alpha}$ and constitute paticular classes 
in the general variety of actions of the latter. 
Their distinguishing feature is the presence of abelian 
translational 
isometries. The description of these actions in terms of 
the superfields 
introduced here makes the isometries manifest and allows 
to construct, 
in the manifestly $(4,4)$ supersymmteric way, the dual
 actions the bosonic 
sectors of which are related to those of the original 
actions via the familiar 
abelian $T$ duality. The dual formulation allows to reveal two 
non-trivial reductions of the relevant bosonic manifold which yield 
most general $4$-dimensional hyper-K\"ahler manifolds 
with one translational 
isometry. We also presented a new sigma model action 
possessing the 
$N=4$, $SU(2)$ superconformal symmetry non-trivially realized  
on harmonic variables. For the off-shell superfield 
action of 
$N=4$ $SU(2)\times U(1)$ WZNW sigma model we found a 
new representation 
in terms of the superfields $q^{2,0}, q^{0,2}$ or 
$Q^{2,0}, Q^{0,2}$ 
on which one additional algebraic constraint is imposed. 

It would be interesting to construct analogous torsionful 
generalizations of 
some other off-shell $(4,4)$ multiplets with finite number 
of auxiliary fields, e.g., of the relaxed tensor multiplet 
\cite{HST}, 
and to find out possible stringy applications of all such 
generalized multiplets.
 
Finally, let us recall that the models considered here admit a 
formulation 
in terms of twisted superfields $q^{1,1}$ and so belong to the 
particular class of torsionful $(4,4)$ sigma models possessing 
mutually 
commuting left and right complex structures on 
the bosonic target \cite{GHR}. It was argued in 
\cite{{EI},{EI1}} that 
the true 
analog of the ``ultimate'' $q^{(+)}$ hypermultiplet in 
the case with torsion 
is the $SU(2)\times SU(2)$ analytic superfield triple 
consisting of 
$q^{1,1}$ and the lagrange multipliers 
$\omega^{1,-1}, \omega^{-1,1}$. One 
can expect that an off-shell formulation of 
general $(4,4)$ sigma models with torsion can be achieved using  
this triple. A generalization of the dual 
$q^{1,1}$ lagrangian 
\p{lagrmul1} was constructed, such that it does not admit 
a formulation 
solely in terms of $q^{1,1}$ and corresponds to a more 
general case with 
non-commuting complex structures. It would be of interest 
to generalize another dual action \p{dualist} along 
similar lines and 
to inquire the relevant target space geometry. 

\vspace{0.5cm}

\noindent{\Large\bf Acknowledgements} 

\vspace{0.3cm}
\noindent 
While this paper was nearly finalized, we learned with 
deepest regret 
about the death of our Teacher and colleague Viktor Isaakovich 
Ogievetsky. 
For many years we benefitted a lot from our collaboration and 
illuminating discussions with him, a brilliant scientist and 
great personality. 
It is painful to realize that nevermore we will have such a 
fortunate opportunity.

We acknowledge a partial support from the INTAS grants 93-127 
and 93-633.
E.I. thanks ENSLAPP, ENS-Lyon, for hospitality extended to him 
during the course of this work.

\end{document}